\documentclass[12pt,preprint]{aastex}

\usepackage{epsf}
\usepackage{graphics}

\newcommand{\kms}{km s$^{-1}$}

\newcommand{\cm}{cm$^{-2}$}

\newcommand{\msun}{$M_{\odot}$}

\newcommand{\chandra}{{\it Chandra }}
\newcommand{\etal}{{\it et al. }}
\shorttitle{z=0 Absorption}
\shortauthors{Mathur et al.}

\begin{document}


\def \charthoffset {\hspace{0.2cm}} \def \charthsep {\hspace{0.3cm}}
\def \chartvsepcap {\vspace{0.3cm}}
\def \chartvsep {\vspace{0.1cm}}
\newcommand{\putchartb}[1]{\clipfig{/home/halley/dgrupe/ps/#1}{85}{20}{0}{275}{192}}
\newcommand{\putchartc}[1]{\clipfig{/home/halley/dgrupe/ps/#1}{55}{33}{19}{275}{195}}
\newcommand{\chartlineb}[2]{\parbox[t]{18cm}{\noindent\charthoffset\putchartb{#1}\charthsep\putchartb{#2}\chartvsep}}

\newcommand{\chartlinec}[2]{\parbox[t]{18cm}{\noindent\charthoffset\putchartc{#1}\charthsep\putchartc{#2}\chartvsep}}

\def\lya{\ifmmode {\rm Ly}\alpha~ \else Ly$\alpha$~\fi}
\def\lyb{\ifmmode {\rm Ly}\beta~ \else Ly$\beta$~\fi}
\def\lyg{\ifmmode {\rm Ly}\gamma~ \else Ly$\gamma$~\fi}
\def\civ{\ifmmode {\rm C}\,{\sc iv}~ \else C\,{\sc iv}~\fi}
\def\civn{\ifmmode {\rm C}\,{\sc iv}~ \else C\,{\sc iv}\fi}
\def\cvn{\ifmmode {\rm C}\,{\sc v}~ \else C\,{\sc v}\fi}
\def\cvin{\ifmmode {\rm C}\,{\sc vi}~ \else C\,{\sc vi}\fi}
\def\nvn{\ifmmode {\rm N}\,{\sc v}~ \else N\,{\sc v}\fi}
\def\nvin{\ifmmode {\rm N}\,{\sc vi}~ \else N\,{\sc vi}\fi}
\def\nviin{\ifmmode {\rm N}\,{\sc vii}~ \else N\,{\sc vii}\fi}
\def\siiii{Si\,{\sc III}~}
\def\oi{{{\rm O}\,\hbox{{\sc i}}~}}
\def\oin{{{\rm O}\,\hbox{{\sc i}}}}
\def\ov{{{\rm O}\,\hbox{{\sc v}}~}}
\def\ovi{{{\rm O}\,\hbox{{\sc vi}}~}}
\def\ovii{{{\rm O}\,\hbox{{\sc vii}}~}}
\def\oviii{{{\rm O}\,\hbox{{\sc viii}}~}}
\def\ovn{{{\rm O}\,\hbox{{\sc v}}}}
\def\ovin{{{\rm O}\,\hbox{{\sc vi}}}}
\def\oviin{{{\rm O}\,\hbox{{\sc vii}}}}
\def\oviiin{{{\rm O}\,\hbox{{\sc viii}}}}
\def\neix{\ifmmode {\rm Ne}\,{\sc ix}~ \else Ne\,{\sc ix}~\fi}
\def\nex{\ifmmode {\rm Ne}\,{\sc x}~ \else Ne\,{\sc x}~\fi}
\def\hi{\ifmmode {\rm H}\,{\sc i}~ \else H\,{\sc i}~\fi}
\def\kms{\rm\,km\,s^{-1}}
\def\hubunits{\rm\,km\,s^{-1}\,Mpc^{-1}}
\def\K{\,{\rm K}}
\def\cm{{\rm cm}}
\def\chandra {{\it Chandra}~}
\def\chandran {{\it Chandra}}

\def\etal   {{\it et~al.}~}
\def\kms {{\rm km}\,{\rm s}^{-1}}


\title{On the nature of the z=0 X-ray absorbers: I. Clues from an
 external group}


\author{Smita Mathur\altaffilmark{1}, Gregory R. Sivakoff\altaffilmark{1},
 Rik J. Williams\altaffilmark{2}, \& Fabrizio Nicastro\altaffilmark{3} }
\altaffiltext{1}
{Astronomy Department, Ohio State University,
    140 W. 18th Ave., Columbus, OH-43210, U.S.A.}
\altaffiltext{2}
{Leiden Observatory, Lieden University, P.O.Box 9513,
2300 RA Leiden, The Netherlands}
\altaffiltext{3}
{SAO, 60 Garden Street, 02138, Cambridge, MA, USA}

\email{smita@astronomy.ohio-state.edu}




\begin{abstract}

Absorption lines of \ovii at redshift zero are observed in high quality
\chandra spectra of extragalactic sightlines. The location of the
absorber producing these lines, whether from the corona of the Galaxy or
from the Local Group or even larger scale structure, has been a matter
of debate. Here we study another poor group like our Local Group to
understand the distribution of column density from galaxy to group
scales. We show that we cannot yet rule out the group origin of z=0
systems. We further argue that the debate over Galactic
vs. extragalactic origin of z=0 systems is premature as they likely
contain both components and predict that future higher resolution
observations will resolve the z=0 systems into multiple components.
\end{abstract}

\keywords{Galaxy: halo -- Local Group -- galaxies: clusters: general --
 galaxies: clusters: individual (NGC~1600) -- intergalactic medium
 --X-rays: galaxies: clusters }

\section{Introduction}
\label{sec|Intro}

 \chandra and {\it XMM--Newton} observations have detected absorption
 lines due to highly ionized elements, notably \ovii, at redshift zero
 toward all extragalactic sight lines with sufficiently high quality
 grating spectra. The location of the absorber producing these lines has
 been a matter of debate, centering primarily on two scenarios: (1) in
 the extended halo of our Galaxy or (2) in the Local Group (LG) or the
 large scale structure around the Galaxy. There are good arguments in
 support of both (1) and (2). Wang et al (2005) find \ovii absorption
 toward the large Magellanic cloud (LMC). Bregman \& Lloyd--Davies
 (2007) find correlated \ovii line strength with soft X-ray emission
 from the Galaxy. Similar observations and theoretical arguments by Fang
 et al. (2006) support scenario (1). Galactic fountain models (Shapiro
 \& Field 1976) also predict a hot halo around star--forming spiral
 galaxies, perhaps including the Galaxy itself. Models of large scale
 structure and galaxy formation, on the other hand, suggest that
 scenario (2) is a plausible source for local absorption. Comparison of
 \ovii absorption strength with the emission measure suggests a
 low-density extended environment (Rasmussen et al. 2003). Direct
 determinations of temperature and density through photo-- and
 collisional--ionization models of the observed absorption line ratios
 do not rule out either scenario (Williams et al. 2005,2006,2007), but
 in at least two cases the velocity dispersion and/or centroid of the
 \ovii\ is inconsistent with known lower--ionization Galactic
 components.

The importance of this debate hinges on the total mass probed by the
\ovii absorption lines, which in turn depends upon the assumed path
length. If the \ovii absorbers are related to the Galaxy, then the
associated mass is insignificant (though it can provide an important
constraint on models of galaxy formation and feedback). On the other
hand, if they trace the large intergalactic scale (or LG) structure, the
associated baryon reservoir becomes comparable to the baryonic mass of
all Local Group galaxies combined. It is well known that a fraction of
LG baryons are ``missing''; the total mass inferred from known galaxies
and in observed cold and hot gas falls short of the dynamical mass of
the LG assuming the baryon fraction is 15\% of the dark matter (Kahn \&
Woltjer 1959; Nicastro et al. 2003 and references therein). It is likely
that these missing baryons are in the not-yet observed warm-hot phase,
which would be traced by \ovii lines.

It is difficult to resolve this issue with simple observations, because
the spectral resolution of current X-ray gratings is insufficient to
resolve the absorption lines into Galactic and LG components; this is
why a variety of approaches were undertaken by different groups cited
above. In fact, there is almost certainly some hot gas associated with
the thick disk of the Galaxy. What we need to know is whether there are
components to the z=0 systems from the Galactic corona, LG, and extended
structures, and if so, what is their relative contribution. In this
article we attempt to answer a simple question: given what is seen in
other poor groups of galaxies, can we rule out
\ovii absorption from intra-group gas at strengths as observed as in
$z=0$ systems? We determine, from the perspective of an observer
centered within NGC 1600 (a poor group of galaxies comparable to the
Local Group), how much the galactic and intragroup media would each
contribute to the local warm-hot column density.

\section{Method}
\label{sec|Method}

The Local Group is a poor group of galaxies, consisting of two main
galaxies and a number of satellite galaxies. Needless to say, there are
other poor groups in the Universe. The physical properties of the
intra-group gas, such as temperature and luminosity, and its radial
distribution are governed by the group gravitational potential, not by
individual galaxies. It would be of interest, therefore, to study the
distribution of gas in and around individual galaxies and in the
intra-group medium in a poor group similar to the Local Group.

One such group, for which \chandra data already exist, is NGC~1600. The
high spatial resolution offered by \chandra is important here, as it
allows exact decomposition of observed intensity profile into the
galactic and group components; this will be clear in the following
discussion. This group is also poor, made of three galaxies, NGC~1600,
NGC~1601, and NGC~1603 of which NGC~1601 is a lenticular galaxy while
the other two are ellipticals. Admittedly, the galaxies in this group
are different from the LG galaxies, but the group richness is similar,
so the properties of the intra-group gas, which are of interest for our
purpose, should be similar, on the order-of-magnitude level, to the
LG. The NGC~1600 group was observed with \chandra in September 2002. The
details of the observations and data reduction are presented in
Sivakoff, Sarazin \& Carlin (2004, hereafter Paper I). While the focus
of their work was on the bright elliptical galaxy NGC~1600, the entire
image of the group was analyzed by these authors (see their Figure 1).

The analysis of the extended X-ray emission from the NGC~1600 group
presented in Paper I is as follows. The X-ray emission
extends all the way from the center of the central galaxy to the
outskirts of the group. From visual inspection of the group
emission, it is clear that there is a smooth transition from the
central galaxy emission to the group emission. The surface brightness
profile of the diffuse X-ray emission is well fit by the sum of two beta
models, each defined as: 
\begin{equation}
I_X(a) = I_0\left[1+\left(\frac{a}{a_c}\right)^2\right]^{-3\beta+1/2}
\end{equation}
where one component corresponds to the interstellar medium of the central
galaxy and the other to the intra-group gas; $I_0$ and $a_c$ are
the central surface brightness and core radius respectively. The group
emission starts to dominate beyond about 25\arcsec\ from the center
corresponding to 7.3 kpc at the source. The emission from the central
galaxy is soft while that from the group is harder. There is also soft
extended emission in/around the other two smaller galaxies. In
addition there is a tail of soft emission connecting NGC~1603 and
NGC~1600 (see Paper I, Figure 7). The center of the group potential lies
to the northeast of the galaxy NGC~1600. The temperature of the group gas
is $\sim 1.5$ keV as deduced from the spectral fits, and consistent with
the temperature of other X-ray bright groups. Note that 1.5 keV
corresponds to a temperature of about $1.6\times 10^7$K; the temperature
range of $10^5$--$10^7$K is called warm-hot, and it is in this
temperature range most of the baryons in the low redshift Universe are
believed to be hiding (Cen \& Ostriker 1999; Nicastro et al. 2004).

From the temperature and the radial surface brightness profile
(described as a double beta model), it is straightforward to estimate
the radial density distribution. In Paper I, the X-ray emitting hot gas
was fitted with an optically thin thermal plasma model (MEKAL model in
XSPEC). In this model, the luminosity is simply $L= P^{\prime} \times
EM$ and $EM$, the emission measure is $\int n_e^2 dV$. The
multiplicative factor $P^{\prime}$ at each wavelength depends on
density, temperature and metal abundances (Mewe et al. 1985). Thus, by
fitting the observed X-ray spectrum in the 0.3--8 keV band, one can
determine the three free parameters of the model, viz. temperature,
density and abundance. In Paper I, the metallicity of the intra-group gas
was not well constrained, but was found to be consistent with
approximately solar (albeit with a factor of $\sim 2$ uncertainty).

For the purpose of this paper, we need to determine the column density
of the warm-hot gas from the perspective of an observer centered on the
NGC~1600 galaxy, which can be calculated by integrating the density 
profile. If we were to observe distant quasars from this vantage point,
 our sight lines would pass
through both the galaxy and group media. Our aim is to determine how much
column density is contributed by each of these components.

%

\section{Results}
\label{result}

In Figure 1, we have plotted the observed density distribution in the
NGC~1600 group. As noted above, the entire surface brightness
distribution is described as a double beta model, one corresponding to
the gas in the galaxy NGC~1600 and one corresponding to the
group. Accordingly, we determine the density distributions of both the
components; the dashed line in figure 1 is for the galaxy NGC~1600, the
dot-dashed line is for the group and the solid line corresponds to the
total. At the distance of the source (D=59.98 Mpc, Prugniel \& Simien
1996), one arcsecond corresponds to 0.29 Kpc. As expected, the density
distribution is dominated by the galactic component in the inner 20
arcseconds, beyond which the group component dominates. In the
transition region around 20 arcseconds from the center, there is a bump
in the density distribution which is an artifact of the fitting
procedure; the exact density around this transition region, however, is
immaterial for the purpose of this paper.

The column density through the gas is simply $N=\int n dr$, which we
have plotted in Figure 2. Again the dashed line corresponds to the
contribution by the galaxy, while the dot-dashed line marks the group
contribution and the solid line is for the total. As expected, the
column density due to the galaxy asymptotically approaches a limit of
$\sim 3\times 10^{20} \cm^{-2}$ at around 10 kpc from the center. The
group contribution is monotonically rising, and dominates the total
column density beyond about 200 kpc. At a distance of 1 Mpc from the
center, the group's column density is about $4.5\times 10^{20}
\cm^{-2}$.

In Figure 3, we have also plotted the mass distributions associated with
the warm-hot gas in the galaxy and the group. As expected, the mass
associated with the galaxy saturates at about $3\times 10^8$\msun. For a
group size of about a Mpc ($\sim 3000^{\prime\prime}$), the mass
associated with the group gas is four orders of magnitude larger, about
$3.5\times 10^{12}$\msun. The solid line in figure 3 corresponds to the
gravitational mass of the group as calculated in Paper I; at 1Mpc this
is $4\times 10^{13}$\msun, implying a baryonic to dark matter ratio of
about 1/10.


\section{Discussion \& Conclusions}
\label{Concl}

As discussed in \S 1, we observe absorption lines of \ovii at $z=0$ in
\chandra spectra of extragalactic objects. The column density of \ovii
varies from about $1.6\times 10^{16} \cm^{-2}$ in Mrk 421 and Mrk 279, to
$6.3\times 10^{15} \cm^{-2}$ in PKS~2155-304, all within a $1\sigma$ range
of $1\times 10^{16} \cm^{-2}$. Assuming a fraction of oxygen in the \ovii
state to be unity (which is the case for the temperatures and densities
observed at $z=0$), and solar abundance to be $\log(n_O/n_H)= -3.13$,
the corresponding total column density is $N_H = 2.2\times 10^{20}
\cm^{-2}$ for [O/H] of $-1$ (a tenth solar metallicity). If the
metallicity is a third solar, the total column density would be $N_H =
6.7\times 10^{19} \cm^{-2}$ and if the metallicity is close to solar,
the corresponding $N_H = 2.2\times 10^{19} \cm^{-2}$. Note that in
groups and clusters of galaxies, metallicities are believed to be of the
order of a third solar. However, the comparison that we make here is
independent of the actual value of the metallicity assumed as long as we
assume it to be the same for the local absorption systems and the group.

In the calculations leading to figures 1, 2, \& 3, solar metallicity was
assumed. As mentioned above, the total column density through the galaxy
NGC~1600 is $N_H = 4\times 10^{20} \cm^{-2}$. Thus, a column density as
observed in the z=0 systems can be easily accommodated in NGC~1600, both
the galaxy and the group. 

There are several caveats to the above statement which we need to
discuss at this stage. (1) Our Galaxy is a spiral galaxy while NGC~1600
is an elliptical, so the distribution of warm/hot gas in the two need
not be similar. (2) The observed, emitting gas in the galaxy and the
group NGC~1600 is too hot to contain significant amounts of \oviin. (3)
The density of the emitting gas is also much higher than what is
observed in z=0 absorption systems and (4) the mass of the NGC~1600
group is an order of magnitude larger than the LG; the observed high
temperature is likely the result of the high mass. We discuss each of
these caveats below.

Elliptical galaxies, such as NGC~1600, are known to be filled with warm
hot gas, resulting in diffuse emission. On the other hand, the X-ray
emission from spiral galaxies is dominated by their point source
population, well correlated with the optical blue light (Palumbo et
al. 1985), while extended diffuse X-ray emission is typically associated
with outflows from extreme starburst activity. The column density of
warm/hot gas through a normal spiral galaxy is thus likely to be much
smaller than that through NGC~1600; for instance, diffuse X-ray emission
has been detected in M31 but it appears to be primarily concentrated in
the bulge and/or associated with hot stars in the disk (Trudolyubov et
al. 2005 and references therein). We chose to analyze the NGC~1600 group
because of its similarity with the Local Group (in terms of being a poor
group), not because of its central galaxy.  Moreover, the question at
hand is not whether the $z=0$ system can arise in the Galaxy (or NGC
1600), but rather {\it could the intergalactic gas in a poor group of
galaxies, such as the Local Group, produce X-ray absorption at the
strength observed at $z=0$?}  For this reason, the exact properties of
the constituent galaxies are immaterial.


For NGC~1600, the column density derived, as shown in Figure 2, is for
the gas emitting in X-rays. Because the emissivity depends strongly on
emission measure, much of the {\it emission} comes from hotter, denser
gas while the z=0 {\it absorption} systems would be dominated by low
density diffuse gas at lower temperature ($\sim 10^6$K). Indeed, at
temperatures above $10^7$K, the fractional ionization of oxygen in \ovii
state is practically zero, so the gas observed in emission would
contribute nil to \ovii absorption (see, e.g., figure 4 in Mathur,
Weinberg \& Chen 2003 for the fractional ionization of oxygen as a
function of temperature).  Examination of Figure 1 in Paper I shows that
the detectable diffuse emission from the NGC~1600 group extends out to
about 70 kpc from the center. At about 300 kpc ($1000^{\prime\prime}$),
the density falls to the value observed in the Mrk~421 z=0 system
($1.2\times 10^{-4} cm^{-3}$; figure 1). At such distances, the gas
could also be cooler, if it hasn't yet fallen into the group
potential. Thus, instead of considering the total column density through
the group, it might be more appropriate to determine the column density
accumulated beyond 300 kpc. From figure 2, this turns out to be $\sim
1\times 10^{20} cm^{-2}$, which can still accommodate the z=0
systems. The assertion of gas being cooler in the outskirts of
groups/clusters appears to be justified: the intracluster gas
temperature of the Coma cluster is 8.21 keV (Hughes et al. 1993). At
such high temperatures, most of the gas is fully ionized, and yet
emission lines of \oviin, \oviiin, and \neix were detected at the
position of the cluster (Takei et al. 2007), presumably from the cooler
outskirts.

It is also interesting to note here that the gas temperature in the
galaxy NGC~1600, even though somewhat lower than that in the group core,
is still quite hot: 0.85 keV ($9.86\times 10^6$K). The fractional
ionization of oxygen in \ovii state at such temperatures is still
practically zero. Thus, the gas seen in emission in the galaxy NGC~1600
will {\it not} produce absorption lines as seen in the z=0 systems.

The total gravitational mass of the Local Group is $\sim 2\times
10^{12}$\msun (Kahn \& Woltjer 1959; Peebles 1995) which is about an
order of magnitude smaller than that of the NGC~1600 group (figure
3). This leads to a column density which is also an order of magnitude
smaller (if the change in mass is a result of the change in central
density), or $N_H = 4\times 10^{19}$\cm, again consistent with what is
observed in the z=0 systems. More importantly, smaller mass of LG would
imply lower temperature than the NGC~1600 group, bringing it closer to
the WHIM temperature range where fractional ionization of \ovii
peaks. Fang et al (2007) detected \oviii absorption line near a small
group containing four galaxies, indicating that the temperature becomes
higher for groups more massive than ours.

Note also that the beta models were fitted to the NGC~1600 group up to a
radius of $180^{\prime\prime}$ and their extrapolation out to
$1000^{\prime\prime}$ may not be appropriate. While an abrupt change in
the density profile is not to be expected, and would not be physical, we
need to keep this caveat in mind. Nonetheless, the analysis of the
density profile of a real poor group like NGC~1600 allows us to make
reasonable estimates of mass and column density profiles, as shown in
figures 2 \& 3. One major argument against the large-scale origin of z=0
systems was made by Collins, Shull \& Giroux (2005) based on, what was
claimed to be, the unacceptably excessive mass of such a structure. They
showed that if the \ovii absorbers are distributed in a spherical shell
centered around the LG at a mean radius of $\approx 1$Mpc, then the
associated total mass would be 12 times larger than the gravitational
mass of the LG (assuming a metallicity of a tenth solar and baryon to
dark matter fraction of 1/6). Such excessive mass results from the
assumption of the shell geometry, and to a lesser extent from the
assumption of metallicity. Williams et al. (2005) have already shown
that the mass can be reasonable for metallicities of a third solar and
if the gas is uniformly distributed over the sphere of $\approx 1$Mpc
radius. With even a more realistic density profile, as shown here by a
beta model for NGC~1600, the baryonic mass lies well below 1/6 of the
dark matter mass as is clear from figure 3 (and the numbers given in
\S3).

Clearly, there are several poor groups in the nearby Universe, and NGC
1600 is just one of them. We will exploit the \chandra archive and
perform similar analysis of other poor groups to characterize their
similarities and differences to draw further inferences on the nature of
the z=0 systems. In particular, it would be of interest to find a group
as massive as the LG and containing spiral galaxies.

 Of course, in studying an external group of galaxies we cannot make
 {\it direct} inferences about the Local Group's matter content, and we
 are thus not claiming a definitive Local Group origin for the $z=0$
 X-ray absorption systems.  However, it appears reasonable to state
 that, given what we observe in poor groups, such an origin cannot be
 ruled out, and is entirely plausible in some circumstances. In fact,
 given the current limitations of X-ray spectrographs, it is perhaps
 premature to have a debate on their Galactic vs. extragalactic origin.
 If we take lessons from other groups, such as NGC~1600, the z=0 systems
 most likely contain both Galactic and Local Group components, and also
 perhaps components from larger structures such as the Local Sheet. We
 predict that future higher resolution observations will resolve the z=0
 systems in two or more components. One would be a low velocity
 component associated with the disk of the Galaxy and one or more may
 arise from extended structures. FUSE, which resolves low- and
 high-velocity components, has a spectral resolution of R=20000 in the
 medium resolution mode. Next generation X-ray spectrographs will have
 to have resolution of at least R=3000 (100 km/s) to begin to resolve
 different velocity components.

%
%

\newpage
\begin{figure*}[t!]
\resizebox{\hsize}{!}{\includegraphics[clip=true]{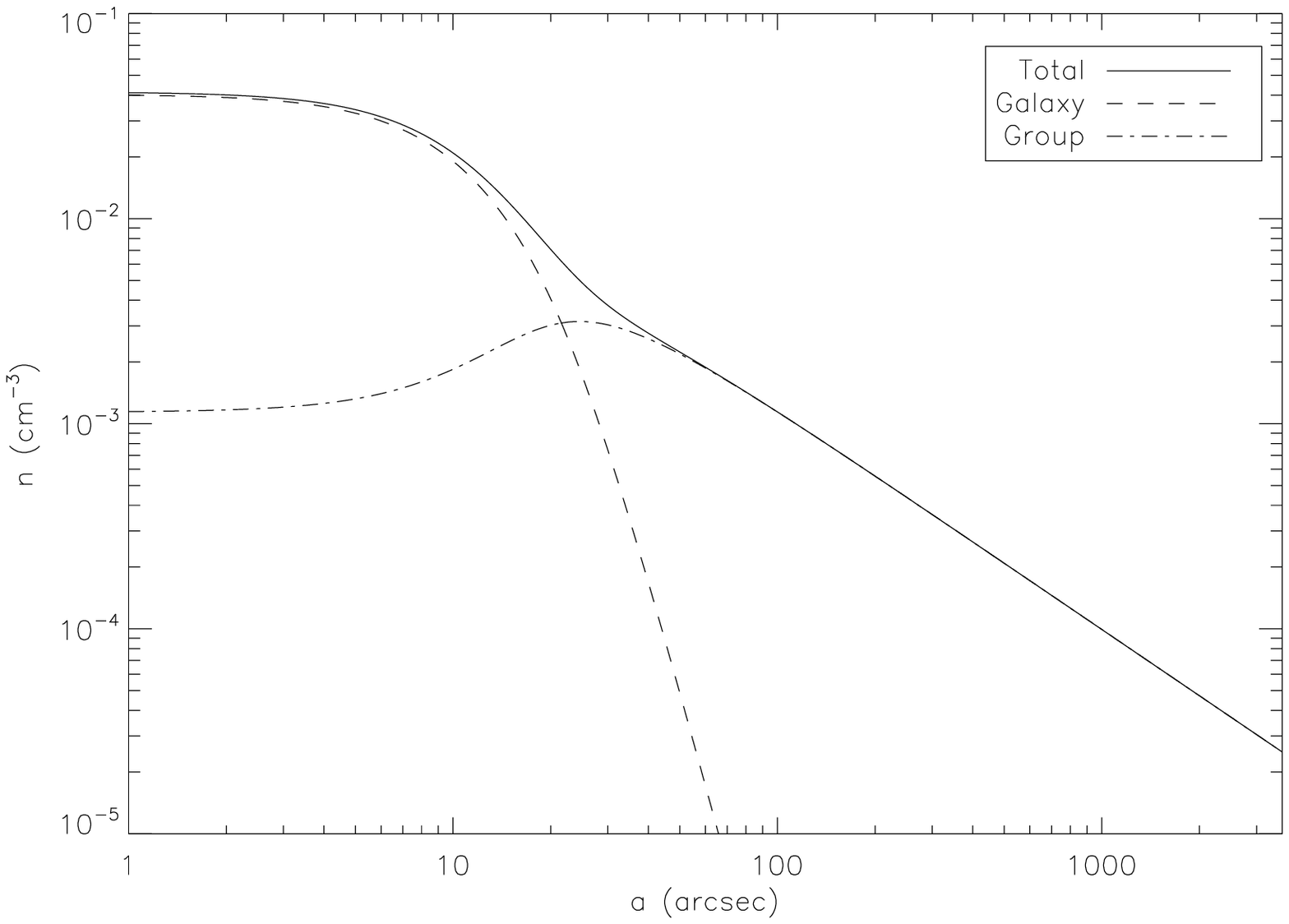}
}
\caption{The gas density profile in the NGC~1600 galaxy (dashed line)
and group (dot-dashed line). The solid line is for the total density
profile.}
\label{fig|density}
\end{figure*}

\begin{figure*}[t!]
\resizebox{\hsize}{!}{\includegraphics[clip=true]{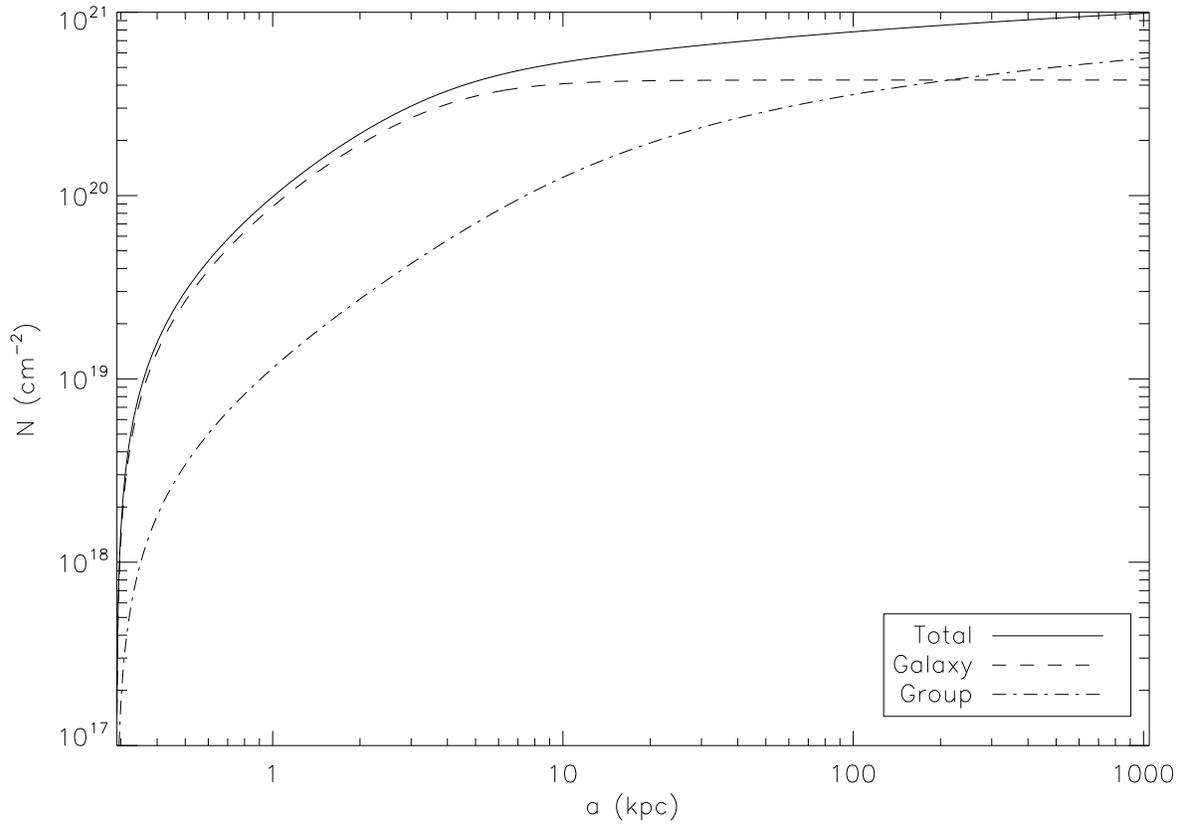}
}
\caption{The column density through NGC~1600. The solid line shows the
 total column density while the dashed and dot-dashed lines mark the
 contributions of the galaxy and the group gas respectively. While the
 galaxy dominates the column density out to 200 kpc, the group column
 density dominates the total for longer sightlines.}
\label{fig|density}
\end{figure*}

\begin{figure*}[t!]
\resizebox{\hsize}{!}{\includegraphics[clip=true]{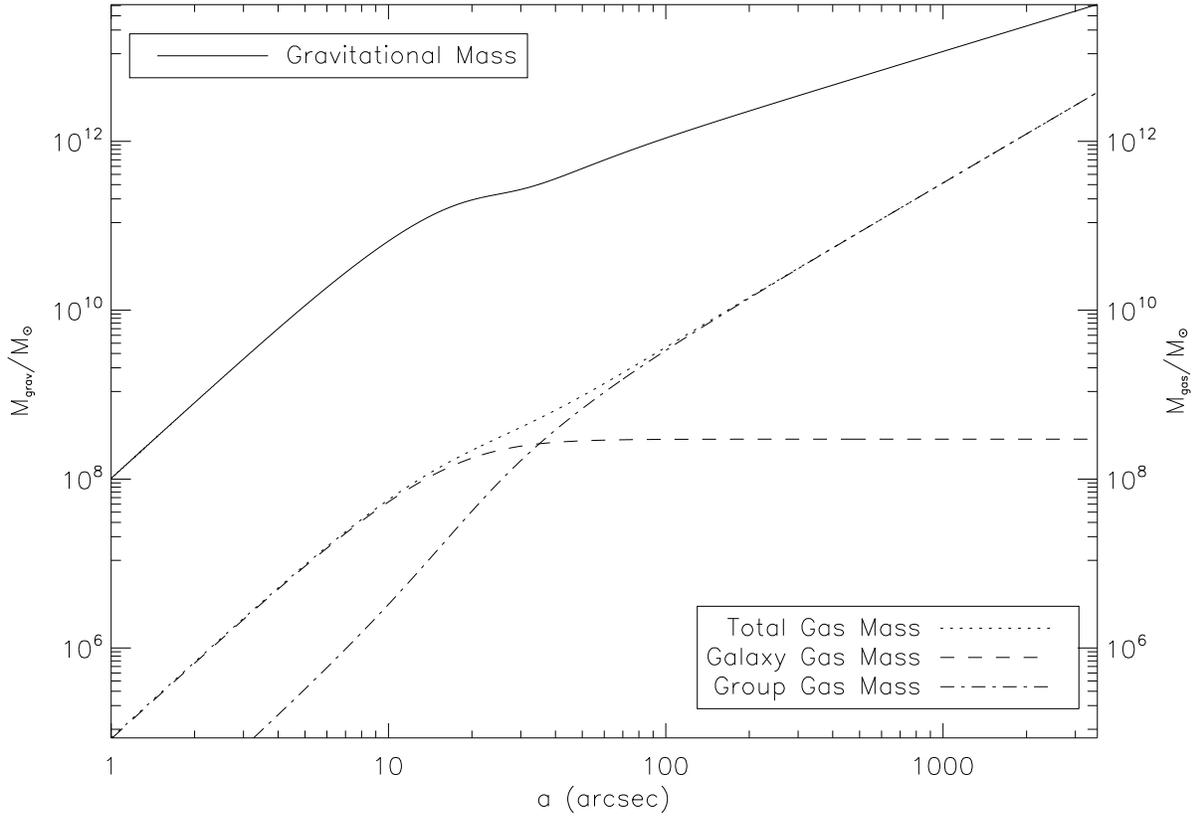}
}
\caption{The gas mass profiles in the NGC~1600 group. The solid line is
 for the total gravitational mass (adapted from Paper I), which is
 significantly larger than the gas mass.}
\label{fig|density}
\end{figure*}

\end{document}